\documentclass[submission,copyright,creativecommons]{eptcs}
\providecommand{\event}{ThEdu 2020} 
\usepackage{breakurl}             

\usepackage{amsmath}
\usepackage{amssymb}
\usepackage{amsfonts}
\usepackage{xcolor}
\usepackage{tikz}
\usetikzlibrary{arrows,shapes}

\newcommand{\myhat}{\ensuremath{\,\hat{\,}}}

\title{Teaching Interactive Proofs to Mathematicians}
\author{Mauricio Ayala-Rinc\'on
\institute{Universidade de Bras\'ilia\\ Bras\'ilia D.F., Brazil}
\email{ayala@unb.br}
\and
Thaynara Arielly de Lima
\institute{Universidade Federal de Goi\'as\\ Goi\^ania, Brazil}
\email{thaynaradelima@ufg.br}
}
\def\titlerunning{Teaching Interactive Proofs to Mathematicians}
\def\authorrunning{M. Ayala-Rinc\'on \& T.A. de Lima}
\begin{document}
\maketitle

\begin{abstract}
This work discusses an approach to teach to mathematicians the importance and effectiveness of the application of Interactive Theorem Proving tools in their specific fields of interest. The approach aims to motivate the use of such tools through short courses. In particular, it is discussed how, using as case-of-study algebraic notions and properties, the use of the proof assistant Prototype Verification System PVS is promoted to interest mathematicians in the development of their mechanized proofs. 
\end{abstract}

\section{Introduction}

Previous academic experiences at our institutions pointed out the relevance of teaching foundations of formal methods to Computer Science (CS) professionals \cite{AR_FMTeaching2019}. In principle, what is expected is that when CS  students learn the foundations of logical deduction in an introductory course in Computational Logic, they also develop technical capabilities to apply logical deduction and inductive methods to verify correctness properties of their programs through the use of proofs assistants.  In the case of the institution of the first author, this is formulated in the teaching methodology for the course on Computational Logic offered to third-semester students in CS, who have a minimal background in programming, discrete mathematics, algebra and calculus \cite{ayala2017applied}. The aim of the course is that CS students develop skills and expertise in this relevant application of mathematical deduction through the performance of verification exercises of correction of simple sorting algorithms (see the NASA PVS library theory {\tt sorting}, developed by members of our group, at \url{https://github.com/nasa/pvslib}).  

In this work, we show how we coined out a completely different
teaching approach to motivate mathematicians to involve in their
efforts in theorem proofs the routine of application of Interactive
Theorem Proving tools (ITPs). The discussion is supported by a PVS
theory used to give short tutorials to mathematicians with this aim
and available at 
\url{http://ayala.mat.unb.br/publications.html}.  The
tutorial deals mainly with notions and theorems of algebra in which
the authors have some expertise (see NASA PVS libraries of Formal
Developments on algebra and groups available at
\url{https://github.com/nasa/pvslib} and~\cite{LGAA2020,SilvaLG18}).

\subsection{Motivation}\label{sec:Motivation}

We aim to establish an adequate teaching approach to present to mathematicians the power of ITPs by showing how these tools can contribute to understanding mathematical theories deeply by analyzing all the particularities and formal requirements to specify and prove mathematical notions and properties.  The target audience is, researchers and graduate students in all areas of mathematics well-trained in pen-and-paper proofs, who are not expected to be truly interested in time-consuming investments in a long training in areas such as Mathematical Logic, Proof Theory, and Logical Deduction, but who may be truly interested in profiting from the benefits of such powerful computational tools.   

Usually, researchers and students from the target audience are capable to use computational tools for numerical and symbolic computations (e.g., Mathematica, Octave, Matlab, Maple) as well as mathematical editors (e.g., {\TeX}, {\LaTeX}), Internet search engines,  and professional communication tools (Zoom, Skype, Teams, etc.), among others, but they are neither confident nor necessarily happy-users of such systems. Therefore,  to involve them in dealing with another kind of computational tool as a proof assistant, the benefits for improving the precision and quality of their professional work (i.e., mathematical proofs) should be made explicit quickly.  

The main benefits of proof assistants should be diligently illustrated to convince such an audience of the importance of following a formal discipline for proving and checking theorems:

\begin{enumerate}
    \item  The importance of specifying consistently mathematical notions respecting the dependency of each notion on previously specified notions, and without allowing any omission; 
    \item  The necessity of fulfilling formally all required cases, avoiding any informal or intuitive argumentation, in the formalization of pen-and-paper proofs;
    \item The importance of using such technologies  to certify the correctness and quality of mathematical proofs. 
\end{enumerate}

With this in mind, an eight-hour tutorial using the proof assistant PVS has been developed. The given material briefly introduces the proof engine of the proof assistant presenting the mathematical background, indeed, Gentzen Sequent Calculus and Induction, just mentioning the existence of associated research fields such as Proof Theory in which proof assistants are based. But the emphasis of the tutorial is on illustrating, considering classical fields of mathematics such as analysis, algebra, and geometry, how an ITP as PVS may be applied to fully specify basic concepts and formalize non necessarily elementary theorems of these fields. 

In this paper we illustrate the proposed tutoring methodology that essentially follows two steps of increasing precision: initially, motivating the application of the ITP to follow the logical reasoning involved in well-known proofs of popular mathematical theorems and then,  focusing on algebra, illustrating how proofs might be fully formalized. All that is done without emphasizing the deductive underground formalisms such as Gentzen Calculus, Higher-Order deduction, and construction of inductive schemata, but just focusing on the technical aspect of the application of the ITP to prove their theorems of interest.

\subsection{A Few Related Works}

Initially, the tutorial provides an overview of PVS by highlighting aspects of its specification language and explaining the Sequent Calculus in which the proof engine of PVS is based, namely, Calculus \`a la Gentzen, besides to present examples where PVS has been successfully used both in academic and industrial environments.  
There is no innovation in such an approach, the emphasis on the target audience has been applied in several courses on formal methods. In addition to our course on Computational Logic for CS students previously mentioned, other approaches to teach formal methods to different audiences have been proposed.  Restricting the discussion to a few and recently reported approaches that are focused on fulfilling quickly the specific interests of the attendants, we can mention the course on Deductive Verification in Why3 by Sandrine Blazy at the \textit{Universit\'e Rennes 1}, to train undergraduate students to develop their own correctness proofs of non-trivial sorting and searching algorithms \cite{Blazy_FMTea2019}; the course  taught at the \textit{\'Ecole Nationale Sup\'erioure d'Informatique pour L'Industrie et L'Entrepise},  as part of the Software Engineering curriculum, by  Catherine Dubois et al,  in which students develop skills on formal methods \cite{Dubois_FMTea2019}; and, the course by Kristin Yvonne Rozier prepared for the Aerospace Engineering departments at Iowa State University and the University of Cincinnati, to train attendants  to look at a verification question and identify what formal methods and tools are applicable to check safety-critical systems \cite{Rozier_FMTea2019}. It is interesting to stress Catherine Dubois et al's position (in \cite{Dubois_FMTea2019}) ratifying the importance of the effectiveness of the teaching approach to meet the specific demands of the target audience (in our case Mathematicians, and their case, CS/Engineers): 

\vspace{2mm}

\begin{minipage}[c]{0.92\linewidth}
\textit{``Students are strongly focused on the direct applicability of the knowledge they are taught, and they are not all going to pursue a professional career in the development of critical systems. Our experience shows that students can gain confidence in formal methods when they understand that, through a rigorous mathematical approach to system specification, they acquire knowledge, skills and abilities that will be useful in their professional future.''}  
\end{minipage}

\vspace{2mm}

Some other considerations that we share, and that although evident are very relevant when learning formal methods,  were also pointed out by Bayer et al in \cite{Bayer2019}, to whom we give special attention.  According to this interesting report by undergraduate students from Jacobs University Bremen about their beginner's experience trying to formalize mathematical theorems: it is relevant to start from knowing the main features of the ITP to be used; having programming experience is not necessarily a plus;  it is relevant to start from knowing the chosen ITP  and what it indeed is;  including ITP in the curricula in fields such as Engineering and Mathematics is not feasible and tutorials may be a better approach to reach the interest of these professionals and researchers; starting from formalizing simple theorems is better than proposing very ambitious tasks; being aware of all details of the pen-and-paper proofs is important to transfer them to the ITP; diminishing the grade of automation at the beginning, is more fruitful than using the full automation power of a proof assistant. 

\subsection{Organization}
Section \ref{sec:motivatingITPs} illustrates how the use of ITPs is motivated through reasoning about classical proofs of popular theorems in different fields of mathematics; Section \ref{sec:classicaldeduction} presents how the relevance of ITPs is illustrated by tutoring full formalizations of notions of group theory and theorems about cyclic, torsion and symmetric groups in the HO predicate calculus; Finally, Section \ref{sec:discussion} discusses the extent to which the authors believe this approach reach the proposed objectives.

\section{Motivating the Use of ITPs}\label{sec:motivatingITPs}

The first step to interest mathematicians in the application of ITPs needs to be very simple but motivating.  Then, our approach proposes the development of deduction exercises that are based almost on axiomatizations allowing mathematicians to follow the reasoning involved in well-known proofs of relevant mathematical results.  For an interesting diversity of mathematicians, it is also important the development of examples from several fields.  In this section, some of such examples are discussed in which almost only predicate deduction and elementar type theory is applied using the proof assistant PVS.

\subsection{Algebra - Propositional Deductions on
  Groups} \label{subsec:algebra}

This example uses just propositional deduction and deals with proofs
based on axioms and properties about symmetric, cyclic, and torsion
groups. These properties are very familiar for advanced undergraduate,
graduate in Mathematics and mathematicians (e.g. see textbooks in
abstract algebra such as~\cite{Fraleigh},~\cite{hungerford80}
or~\cite{herstein75}).  The objective of such an example is just
introducing mathematicians to the elements of the ITP tool.  Using
such an example, we explain the environments of the proof assistant,
in which specifications are given and proofs are executed. This
includes a quick start on the syntax of the specification language,
basic types, and the mechanics of the deductive engine, which for the
case of PVS are a functional language and sequent calculus,
respectively.

Propositional variables  for {\tt Cyclic, Abelian, Symmetric}, and finite degree ({\tt FiniteDegree}) groups are specified. Also, a propositional variable for symmetric groups of degree greater than two is given, {\tt nGT2}.
Cyclic groups contain a generator $y$ such that for each element $x$ in the group,  there exists an integer  $i$ such that $x=y^i$; in Abelian groups the binary group operation, $*$, is commutative. The symmetric group of degree $n$ is given by the set of $n$-permutations (bijective functions on the set $\{1, \ldots,n\}$) with neutral element given by the identity function and binary operator given by functional composition.  A group is said to be of finite degree if for all $x$ in the group there exists a natural $n$ such that $x^n$ is the neutral element.   
Indeed, the granularity of a group as a structure $G= \langle  S, *, e\rangle$, where $S$ is a non-empty set with a binary associative function $*$ with  neutral element $e$, and such that $*$ allows inverses (i.e., for all $x\in S$ there exists $y$ such that  $x*y = y*x = e$),  is not specified in this example but will be given in the next step of the tutoring approach, where full formalizations are provided. 

Initially, some axioms specifying well-known properties are given. 

\begin{verbatim}
  Ax01 : AXIOM Cyclic => Abelian
  Ax02 : AXIOM Symmetric_n AND nGT2 => NOT Abelian
  Ax03 : AXIOM FiniteDegree <=>  Torsion
  Ax04 : AXIOM Finite => FiniteDegree
  Ax05 : AXIOM NOT Finite <=> Infinite
  Ax06 : AXIOM Abelian AND Torsion <=> AbelTorsion
\end{verbatim}

And after that, simple propositional (well-know)  consequences of these axioms are proposed to present the specification and proof environments of the ITP.  For instance, the first conjecture below is a consequence of the first and second axiom; the second, of the third and fourth axioms, and so on.

\begin{verbatim}
  Pr01  : CONJECTURE Symmetric_n AND Cyclic => NOT nGT2   
  Pr02 :  CONJECTURE Finite => Torsion                    
  Pr03 :  CONJECTURE Cyclic AND Finite => AbelTorsion     
  Pr04 :  CONJECTURE NOT Torsion => Infinite    
\end{verbatim}

In particular, in PVS, {\tt AXIOMS} may be also called {\tt CONJECTURES} that may be both promoted to {\tt LEMMAS} once proofs are completed. 

The discussion at this point includes providing information about the existence and importance of the field of proof theory and about the relation between PVS propositional proof commands and Gentzen's sequent rules; namely, PVS proof commands as {\tt (flatten)} and {\tt (split)} versus Gentzen rules as $(L_\wedge), (R_\vee)$ and $(R_\to)$ and,    $(R_\wedge), (L_\vee)$ and $(L_\to)$, respectively (see e.g., Chapter 4 in \cite{ayala2017applied} or slides that accompany the tutorial ``Interactive Proving Mathematical Theorems'' available at \url{http://ayala.mat.unb.br/Summer_UnB_2020}). Also, at this point, it is relevant to explain how axioms, lemmas, and conjectures can be invoked or charged as antecedent formulas of the goal sequent using PVS commands as {\tt (lemma) and (rewrite)}.

Besides, it is important to clarify to the attendants the limitations of the propositional language used in such an example explaining that the expressiveness of the language and logical system of the ITP would be enough to provide precise specifications of all required notions and properties and to complete all holes in the proofs.   For doing that, additional elements in the language such as quantifiers may be discussed letting clear how one could, for example, specify (and formalize) with precision the axiom {\tt Ax01} above:

\[\begin{array}{ccc}\underbrace{\exists y \in G: \forall x\in G: \exists i\in \mathbb{Z} : x = y^i }  & \Rightarrow & \underbrace{\forall u, v\in G: u*v = v*u}\\
{\tt Cyclic} && {\tt Abelian}
\end{array}
\]

Also, it would be of interest to discuss elements such as induction that will be required to formalize the  intuitive and simple  pencil-and-paper proof of this property; essentially, the observation below.

\[ \exists i, j : u = y^j, v = y^i \Rightarrow y^i * y^j = y^{i+j} = y^{j+i} = y^j * y^i \]

\subsection{Analysis - a Topological Proof of Infinitude of Primes}

Probably, the most well-known proof of the infinitude of primes is the one presented in \emph{Elements IX} and attributed to Euclid, but here we chose an elegant proof about this fact by using a topological argumentation due to F\"{u}rstenberg \cite{Fuerstenberg1955}.  In the following, we will explain F\"{u}rstenberg's proof and how his lines of reasoning may be specified and formalized to lead the attendants to easily conclude. A detailed and readable explanation of such proof is available in Chapter 1 of \cite{theBook2018}.  

Firstly,  we consider sets of the form $\{ a + n \cdot b \;|\; n \in \mathbb{Z} \}$, where  $a$ and $b$ are integer numbers with $b > 0$. The elements of each such set provide an infinite arithmetic progression from $a$ with period $b$. Sets that may be given in such a form satisfy the predicate {\tt add\_cyclic?}.  In PVS,  {\tt add\_cyclic?}$(N)$ if and only if  $N$ is of type {\tt set[int]} and $N = \{ a + n \cdot b \;|\; n : {\tt int}   \}$, for $a$ and $b$ of respective types {\tt int} and {\tt posnat}.

A topology consisting of \textit{open} sets that are (possibly infinite) unions of such integer sets is given: {\tt open?}$(X)$ if and only if $X$ is the union of sets that satisfy the predicate {\tt add\_cyclic?}, given by the predicate {\tt Union\_add\_cyclics?}. In this topology, any non-empty open set is infinite, and complements of open sets are \textit{closed}, and vice versa, as expected.   

Also, finite intersections of open sets are open (i.e., {\tt fin\_int\_open?}$(X) \Rightarrow$ {\tt open?}$(X)$) as finite unions of closed sets are closed (i.e., {\tt fin\_union\_closed?}$(X) \Rightarrow$ {\tt closed?}$(X)$). 
Notice that a set of the form $\{ a + n \cdot b \;|\; n \in \mathbb{Z} \}$ is  also closed since its complement is the (finite) union of {\tt add\_cyclic?} sets as below.
\[\displaystyle\bigcup_{j=1}^{b-1} \{ (a+j) + n\cdot b \;|\; n \in \mathbb{Z} \} \]

Now, consider the set {\tt per\_PRIMES} given by the union of all {\tt add\_cyclic?} sets of the form  $\{0 + n \cdot p \;|\; n \in \mathbb{Z}\}$, where $p$ in {\tt PRIMES}. Notice that the complement of {\tt per\_PRIMES} is the set  $\{-1, 1\}$. 

If {\tt PRIMES} were finite, as a finite union of closed sets,  {\tt per\_PRIMES} would be closed too, and consequently, its complement $\{-1, 1\}$ would be open, which contradicts the fact that any non-empty open set is infinite.  Thus, {\tt PRIMES} is infinite.

Below, we include a PVS specification of this proof that is mainly based on axioms ({\tt Ax00} - {\tt Ax09}), letting to the mathematicians the task to prove five easy conjectures to complete F\"{u}rstenberg's proof.

\begin{verbatim}
 add_cyclic?, open?, Union_add_cyclics?, fin_int_open?, fin_union_closed? 
    : pred[set[int]]

 PRIMES : set[int]
 X,Y : VAR set[int]

 finite?: pred[set[int]] = is_finite[int]

 Ax00: AXIOM add_cyclic?(emptyset)
 Ax01: AXIOM open?(X) <=> Union_add_cyclics?(X)
 Ax02: AXIOM Union_add_cyclics?(X) AND Union_add_cyclics?(Y) =>
       Union_add_cyclics?(union(X,Y))
 Prop01: CONJECTURE open?(X) AND open?(Y) => open?(union(X,Y))
 Ax03: AXIOM add_cyclic?(X) => Union_add_cyclics?(X)
 Ax04: AXIOM fin_int_open?(X) => Union_add_cyclics?(X)
 Prop02: CONJECTURE fin_int_open?(X) => open?(X) 
 
 closed?(X): bool = open?(complement(X))

 Ax05: AXIOM X /= emptyset AND Union_add_cyclics?(X) => NOT finite?(X)
 Ax06: AXIOM add_cyclic?(X) => 
       EXISTS (Y:set[int]): Union_add_cyclics?(Y) AND X = complement(Y)
 Prop03: CONJECTURE add_cyclic?(X) => closed?(X)
 Ax07: AXIOM fin_union_closed?(X) => closed?(X)
 per_PRIMES: set[int] = complement({x: int | x = 1 OR x = -1})
 Ax08: AXIOM Union_add_cyclics?(per_PRIMES) 
 Ax09: AXIOM finite?(PRIMES) => fin_union_closed?(per_PRIMES)
 Prop04: CONJECTURE finite?({x: int | x = 1 OR x = -1})
 Prop05: CONJECTURE NOT finite?(PRIMES)
\end{verbatim}

In this theory we define open sets restricted to the topology to be considered in the proof, as one can see in specification {\tt Ax01}. However, we point out that {\tt Prop01}, {\tt Prop02} and {\tt Ax07} hold for any topology.

The specification uses PVS prelude set objects:  {\tt emptyset}, {\tt is\_finite}, and {\tt complement} and a minimum of PVS prelude formulas on sets. 

As an illustration the formalization of conjecture {\tt Prop05}
consists just of application of axioms {\tt Ax07} and {\tt Ax09} to
conclude {\tt closed?(per\_PRIMES)}. From that, one has that {\tt
  open?}$(\{-1, 1\})$ using a PVS prelude formula that states the
nilpotency of the complement operator on sets. By {\tt Ax01} and {\tt
  Ax05} one obtains that this set is infinite, which using {\tt
  Prop04} gives the desired contradiction.  Below, a few additional
details on how this is done in PVS are provided. The full proof is
part of the model
tutorial\footnote{\url{http://ayala.mat.unb.br/publications.html}}
that accompanies this paper.

PVS separates the specification from the formalization in files with extensions ``pvs'' and ``prf'', respectively. Goals in PVS are sequents. For proving {\tt Prop05}, once the proof environment is open, the goal 
{\tt finite?(PRIMES) $\vdash$} appears. 

The PVS proof command {\tt (lemma Ax09)} is then used to charge this axiom, and by propositional simplification the formula {\tt fin\_union\_closed?(per\_PRIMES)} becomes an antecedent of the goal sequent. Then, axiom {\tt Ax07} is invoked and instantiated with {\tt per\_PRIMES} (using the PVS proof command {\tt (inst ``pre\_PRIMES'')}), obtaining, by propositional simplification, the sequent:

{\tt closed?(per\_PRIMES), fin\_union\_closed?(per\_PRIMES), finite?(PRIMES) $\vdash$}

Expanding the definition of {\tt closed?}, the formula {\tt closed?(per\_PRIMES)} becomes the antecedent formula {\tt  open?(complement(per\_PRIMES))}.  And then, expanding {\tt per\_PRIMES} one obtains the sequent:

{\tt  open?(complement(complement(\{1,-1\}))),

      fin\_union\_closed?(complement(\{1,-1\})),
      finite?(PRIMES) $\vdash$}

By applying a PVS prelude lemma on nilpotency of the operator {\tt complement} on sets, the first formula becomes {\tt open?(\{1,-1\})}, and applying {\tt Ax01} one obtains the sequent:

{\tt Union\_add\_cyclics?(\{1,-1\}),

     open?(\{1,-1\}),
     fin\_union\_closed?(complement(\{1,-1\})),
     finite?(PRIMES) $\vdash$}
     
Invoking and  instantiating adequately {\tt Ax05} and then, by propositional simplification, one obtains the sequent:

{\tt \{1,-1\} $\neq\emptyset \Rightarrow$ 
       NOT finite?(\{1,-1\}),
     Union\_add\_cyclics?(\{1,-1\}),
     
     open?(\{1,-1\}),
     fin\_union\_closed?(complement(\{1,-1\})),
     finite?(PRIMES) $\vdash$}

The proof is then concluded, by application of {\tt Prop04} that gives the formula {\tt finite?(\{1,-1\})} and expanding the definition of the empty set, $\emptyset$: {\tt \{x:int : FALSE\}}. Thus, {\tt \{1,-1\} $\neq$ \{x:int | FALSE\}} is proved by decomposing the equality, using the PVS command {\tt (decompose-equality)} that gives the inconsistent antecedent formula  {\tt FORALL(x : int): NOT (x = 1 OR x = -1)}. One concludes, instantiating this formula either with  {\tt 1} or with {\tt -1}.

For teaching issues, this theory does not consider algebraic aspects of sets of integers, but all the required aspects may be considered to obtain a complete formalization (as those in \cite{BaazHLRS08} and \cite{Eberl20a}, obtained using CERES and Isabelle, respectively) as a nice further exercise.

\subsection{Geometry - Pick's Theorem}  

\begin{figure}[h]
\centering
    \begin{tikzpicture}[-,scale=1.8, auto,swap]
        \node (a) at (0,0) {$\bullet$};
        \node (b) at (0,1) {$\bullet$};
        \node (c) at (1,1) {$\bullet$};
        \node (d) at (1,0) {$\bullet$};

        \path (a) edge node {} (b);
       \path (a) edge node {} (c);
        \path (b) edge node {} (c);
        \path (c) edge node {} (d);
        \path (d) edge node {} (a);
    \end{tikzpicture}
    \caption{Euler's formula for three faces (the external face counts), four vertices and five edges}
    \label{fig:EulerFormula}
\end{figure}
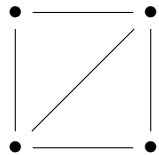

Euler's formula  states that if $G$ is a connected plane graph with $n$ vertices, $e$ edges and $f$ faces, then (see the example in Fig. \ref{fig:EulerFormula}):

\[ n - e + f = 2 \]

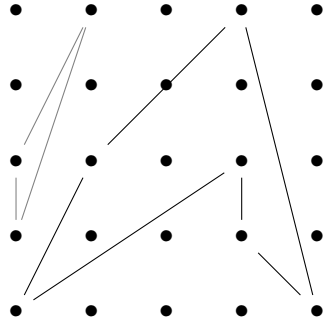
\begin{figure}[h]
\centering
    \begin{tikzpicture}[-,scale=1, auto,swap]
        \node (a) at (0,0) {$\bullet$};
        \node (b) at (0,1) {$\bullet$};
        \node (c) at (0,2) {$\bullet$};
        \node (d) at (0,3) {$\bullet$};
        \node (e) at (0,4) {$\bullet$};
        \node (f) at (1,0) {$\bullet$};
        \node (g) at (1,1) {$\bullet$};
        \node (h) at (1,2) {$\bullet$};
        \node (i) at (1,3) {$\bullet$};
        \node (j) at (1,4) {$\bullet$};
        \node (k) at (2,0) {$\bullet$};
        \node (l) at (2,1) {$\bullet$};
        \node (m) at (2,2) {$\bullet$};
        \node (n) at (2,3) {$\bullet$};
        \node (o) at (2,4) {$\bullet$};
        \node (p) at (3,0) {$\bullet$};
        \node (q) at (3,1) {$\bullet$};
        \node (r) at (3,2) {$\bullet$};
        \node (s) at (3,3) {$\bullet$};
        \node (t) at (3,4) {$\bullet$};
        \node (u) at (4,0) {$\bullet$};
        \node (v) at (4,1) {$\bullet$};
        \node (w) at (4,2) {$\bullet$};
        \node (x) at (4,3) {$\bullet$};
        \node (y) at (4,4) {$\bullet$};

        \path (a) edge node {} (r);
        \path (r) edge node {} (q);
        \path (q) edge node {} (u);
        \path (u) edge node {} (t);
        \path (t) edge node {} (h);
        \path (h) edge node {} (a);

        \path[gray] (b) edge node {} (j);
        \path[gray] (j) edge node {} (c);
        \path[gray] (c) edge node {} (b);
        
    \end{tikzpicture}
    \caption{An elementary Polygon (gray) and an integral Polygon (black)  - with vertices in $\mathbb{Z}^2$ }
    \label{fig:IntPoly}
\end{figure}

Pick's theorem from 1899 is a classical consequence from Euler's formula that establishes another formula for the area of \textit{integral polygons} in $\mathbb{R}^2$, where the vertices of such integral polygons belong to the lattice $\mathbb{Z}^2$.  See the example in Fig. \ref{fig:IntPoly}.  

\begin{figure}[h]
\centering
    \begin{tikzpicture}[-,scale=1, auto,swap]
        \node (a) at (0,0) {$\bullet$};
        \node (b) at (0,1) {$\bullet$};
        \node (c) at (0,2) {$\bullet$};
        \node (d) at (0,3) {$\bullet$};
        \node (e) at (0,4) {$\bullet$};
        \node (f) at (1,0) {$\bullet$};
        \node (g) at (1,1) {$\bullet$};
        \node (h) at (1,2) {$\bullet$};
        \node (i) at (1,3) {$\bullet$};
        \node (j) at (1,4) {$\bullet$};
        \node (k) at (2,0) {$\bullet$};
        \node (l) at (2,1) {$\bullet$};
        \node (m) at (2,2) {$\bullet$};
        \node (n) at (2,3) {$\bullet$};
        \node (o) at (2,4) {$\bullet$};
        \node (p) at (3,0) {$\bullet$};
        \node (q) at (3,1) {$\bullet$};
        \node (r) at (3,2) {$\bullet$};
        \node (s) at (3,3) {$\bullet$};
        \node (t) at (3,4) {$\bullet$};
        \node (u) at (4,0) {$\bullet$};
        \node (v) at (4,1) {$\bullet$};
        \node (w) at (4,2) {$\bullet$};
        \node (x) at (4,3) {$\bullet$};
        \node (y) at (4,4) {$\bullet$};

        \path (a) edge node {} (r);
        \path (r) edge node {} (q);
        \path (q) edge node {} (u);
        \path (u) edge node {} (t);
        \path (t) edge node {} (h);
        \path (h) edge node {} (a);
        \path (a) edge node {} (g);
        \path (g) edge node {} (r);
        \path (g) edge node {} (h);
        \path (h) edge node {} (m);
        \path (m) edge node {} (n);
        \path (n) edge node {} (s);
        \path (s) edge node {} (t);
        \path (u) edge node {} (r);
        \path (u) edge node {} (s);
        \path (g) edge node {} (m);
        \path (m) edge node {} (r);
        \path (r) edge node {} (s);
        \path (m) edge node {} (s);
        
    \end{tikzpicture}
    \caption{Triangulation of an Integral Polygon in Elementary Polygons }
    \label{fig:Triangulation}
\end{figure}
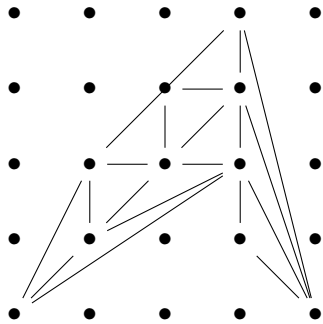

An \textit{elementary} polygon is a convex integral polygon that intersects only its vertices in the lattice $\mathbb{Z}^2$. See an example also in Fig. \ref{fig:IntPoly}.  Elementary polygons that are triangles have an area equal to $1/2$.  Thus, the area of any integral polygon is the half of the number of elementary triangles inside the polygon, for any triangulation of the polygon in elementary triangles. See an example in Fig. \ref{fig:Triangulation}.

Pick's theorem sees an elementary triangulation of an integral polygon as a plane graph and applies Euler's formula to express its area in terms of its vertices.

 Let {\tt f} be the number of faces in an elementary triangulation of an integral polygon {\tt P}, {\tt n = ni + nb} be the number of vertices in the polygon that is equal to the sum of internal and boundary vertices, {\tt e = ei + eb} be the number of edges  in the polygon that is equal to the sum of internal and boundary edges.  First, notice that the number of boundary vertices and edges is the same. Also, notice that each elementary triangle in the boundary has a boundary and two internal  edges, and any internal elementary triangle has three internal edges. These simple observations are given as axioms {\tt Ax01, Ax02, Ax03} and {\tt Ax04} in the PVS specification below, where {\tt ni, nb, n, ei, eb} and {\tt  e}  are of type natural ({\tt nat}), {\tt f} of type natural positive ({\tt posnat)} and, {\tt Area\_p} of type real ({\tt real}).  In particular, the left-hand side of {\tt Ax04} counts the number of edges in each of the {\tt f - 1} elementary triangles, but each internal edge is counted twice.   

\begin{verbatim}
Ax01 : AXIOM ni + nb = n
Ax02 : AXIOM ei + eb = e
Ax03 : AXIOM eb = nb
Ax04 : AXIOM 3 * (f - 1) = 2 * ei + eb
EulerFormula : AXIOM n - e + f = 2
AreaIntPoly : AXIOM Area_p = (f - 1) / 2  
\end{verbatim} 

Pick's theorem establishes the area of any integral polygon {\tt P} in terms of its internal and boundary vertices by the formula:

\[{\tt Area\_P = ni + nb/2 - 1}\]

Euler's formula and the area of an integral polygon, {\tt Area\_P}, are thus given as axioms.  The second is given as the sum of the area of the {\tt f - 1 } elementary triangles in the triangulation of the polygon. 

Below, {\tt Prop01} is a simple reformulation of {\tt Ax04}, and {\tt Prop02} is a consequence of {\tt Prop 01} and {\tt Ax02}.  {\tt Prop03} is a reformulation of {\tt Prop02}.  Finally, Pick's theorem is obtained from {\tt Prop02}, (a simple variant of) the Euler's formula, {\tt Ax01} and {\tt Ax03}, and the previous axiom for the area of integral  polygons. 
\begin{verbatim}
Prop01 : CONJECTURE 3 * f = 2 * ei + 2 * eb - eb + 3
Prop02 : CONJECTURE 3 * f = 2 * e - eb + 3
Prop03 : CONJECTURE f = 2 * (e - f) - eb + 3
EulerFormula_variant : CONJECTURE e - f = n - 2  

Pick_Theorem : CONJECTURE Area_p = ni + nb / 2 - 1
\end{verbatim}
 
An example such this one allows introducing attendants the algebraic deductive engine of the proof assistant illustrating the grade of numerical automation of the system.  Of course, a complete formalization of this proof (as the one reported in \cite{Harrison11} using HOL Light) is an interesting challenge for the attendants.

\section{Teaching \emph{Mathematical} Deduction}\label{sec:classicaldeduction}

After presenting examples such as the ones given in the previous section, the second step of the tutoring approach aims to teach the real usability of ITPs by illustrating how mathematical notions can be specified with precision and how elaborated proofs can be completely formalized. For this proposal, we illustrate here how we do this with algebraic notions and properties. Also, we remark the importance of providing information about important mathematical formal developments.  

\subsection{Complete Formalizations About Cyclic and Torsion Groups} \label{subsec:simple_deduction}
From Subsection \ref{subsec:algebra}, that brings classical theorems about groups from abstract algebra, it is expected that the attendants had realized that it is necessary a more powerful language (than the one of propositional logic) to reach a full formalization of such mathematical theorems. For example, the specification of the group, $\langle S, *, e\rangle$, itself requires the language of predicate logic to quantify elements of the group to specify, for instance, closure of the binary operator $*$: \textit{for all $x,y$ in $S$, $x*y$ belongs to $S$}.
 
 In the PVS theories {\tt pred\_algebra} and {\tt symmetric\_n}, (part of the tutorial cited in the introduction) we present a full formalization of axioms and conjectures presented in Subsection \ref{subsec:algebra}.  In the tutorial,  the adequate specification of the required definitions and properties are discussed, and attendants are encouraged to prove the proposed theorems following a progressively increasing level of difficulty. By illustrating fully formalizations of mathematical objects that are part of their professional reality  attendants are not only motivated to apply ITPs in their research but also to reasoning more thoroughly about issues and gaps that frequently happen in their pen-and-paper proofs.
 
Besides motivating and promoting the increase of skills in formalization of mathematics, this methodology also enables attendants to realize the effectiveness of ITPs as important tools to document mathematics, certify proofs, and check their correctness. All that naturally contributes to a better understanding of mathematical theories, since by using an ITP it is impossible to omit details of the proof or to reach the conclusion of a theorem omitting hypothesis. This contrasts with the pen-and-paper proving approach that usually accepts results omitting \textit{simple} but \textit{crucial} steps that are accepted as ``trivial'' or as ``analogously'' provable.  

Even though the authors participated in the development of several elaborated results in algebra, the theorems included in the tutorial were calibrated in such a manner that they bring simple but challenging formalizations exercises to the attendants. The tutorial includes specification of groups, subgroups, and cyclic, torsion and symmetric groups, and basic properties such as unicity of the identity of groups, left- and right-cancellative laws, the inverse of the inverse equals identity, and the inverse of multiplications equals swapped multiplication of the inverses, among others.  Also, more elaborated results such as cyclic groups are Abelian, finite groups are torsion groups, finiteness of symmetric groups,  the set of natural exponents of an element of a finite group is a subgroup, etc.

In the following, we depict the formalization of the property \emph{Every cyclic group is Abelian} to exemplify how more complex mathematical formalizations are introduced along with the tutorial.

Firstly, in the theory {\tt pred\_algebra} which contains as parameters a  non-interpreted non-empty type {\tt T}, a binary operator {\tt *:[T,T -> T]} and a constant {\tt e} of type {\tt T}, we specify the structure of groups and Abelian groups as a predicate. 

\begin{verbatim}
G : VAR set[T]    % G is a variable of type set of elements of T
                  % (G) would be the type of elements of G, subtype of T
                  
closed?(G): bool = FORALL (x,y:(G)): member(x*y,G)      

inv_exists?(G): bool = FORALL (x:(G)): EXISTS (y:(G)): x * y = e AND y * x = e

group?(G): bool = closed?(G) AND associative?[(G)](*) AND member(e,G) AND
                  identity?[(G)](*)(e) AND inv_exists?(G)
                  
abelian_group?(G): bool = group?(G) AND commutative?[(G)](*)
\end{verbatim}

Such specification is naturally explained to the attendants without much justification about the logical system required (HO predicate logic). Some of the used predicates are included in the prelude of PVS, for instance, {\tt associative?}, {\tt commutative?} and {\tt identity?}. For the first two, it is necessary to provide arguments about the type {\tt (G)} over which the operator {\tt *} is associative or commutative, and for the third one, additionally, the argument {\tt e} that acts as a neutral element for the operator {\tt *} over {\tt (G)}.  

Also, we introduce the specification of the inverse of an element in a group, {\tt inv} by using the PVS indefinite operator  {\tt choose} that is well-defined over non-empty sets. 

\begin{verbatim}
inv(G:(group?))(x:(G)): (G) = choose({y:(G) | x * y = e AND y * x = e})
\end{verbatim}

At this point, we have the opportunity to enrich the discussion on types explaining the \textit{type correctness conditions} (TCCs) that are generated by PVS along the process of type checking. The PVS type checker generates for the function {\tt inv} the TCC  {\tt nonempty?(\{y:(G) | x * y = e AND y * x = e\})}.  This TCC should be proved manually using the type of {\tt G} and expanding the definition of  {\tt group?}.  In PVS TCCs are proof obligations that when cannot be automatically proved, the user should discharge to obtain {\em complete} formalizations.   

In order to define cyclic groups, we specify the exponent of an element {\tt y} of {\tt T} to a natural {\tt n} regarding the operator {\tt *}, as a recursive function over the type of natural numbers ({\tt nat}) and, then, we extend such function over integer numbers for elements of type {\tt (G)} using the inverse of {\tt y}. 

\begin{verbatim}
^(y : T, n : nat ) : RECURSIVE T =
      	       	      IF n = 0 THEN e ELSE  y * ^(y, n-1) ENDIF
                     MEASURE n

^(G:(group?))(y:(G), i: int) : T = IF i >= 0 THEN ^(y,i)
                                    ELSE ^(inv(G)(y),-i) ENDIF
\end{verbatim}

 In PVS the specifier must provide a decreasing {\tt MEASURE} to define each recursive function, and this measure, stated in functions of the parameters of the function,  should decrease along chains of successive recursive calls. For the case of  {\tt \myhat(y,n)} above, the measure {\tt n} is the simplest adequate choice. When the type checker is executed, it provides the \textit{termination} TCC  {\tt caret\_TCC2} below, whose proof ensures the termination of such function. 

\begin{verbatim}
caret_TCC2: OBLIGATION FORALL (n: nat): NOT n = 0 IMPLIES n - 1 < n
\end{verbatim}


Finally, to extend the notion of exponent from naturals to integers (that is only possible over groups), the operator {\tt \myhat(G)(y:(G),i: int)} is specified as {\tt \myhat(y,i)} if $i$ is a natural number, and otherwise as {\tt \myhat(inv(G)(y), -i)}.

The next step is to specify the notion of a cyclic group and then the property that states that every cyclic group is Abelian. 

\begin{verbatim}
 cyclic?(G): bool = group?(G) AND 
                    EXISTS (y: (G)): 
                       FORALL(x:(G)): EXISTS (n: int): ^(G)(y,n) = x
\end{verbatim}

In order to prove that cyclic groups are Abelian, accordingly to the usual mathematical intuition about why this property holds, as mentioned in the end of Subsection \ref{subsec:algebra} (i.e., any pair of elements  $u,v$ of a cyclic group are such that if $y$ is the generator of the group  then for some integers $i$ and $j$, 
$ u * v = y ^ i * y^ j = y ^ {i + j} = y ^ {j + i} =y^j* y^i$), we ask the attendants to assume the \textit{conjectures} {\tt power\_add\_aux} and {\tt power\_add\_aux2}  presented below, and apply them to establish by case analysis on the sign of $i$ and $j$ the required law for integer exponents, as given by the conjecture {\tt power\_add}. PVS allows us to specify and prove a lemma using previously specified but unproved ``conjectures''. However, to provide a full formalization of a lemma all proofs of the previously used conjectures as well as all TCCs must be completed.  The recommended discipline is to specify any unproved lemma as a ``{\tt CONJECTURE}'' and only after it is proved, as a ``{\tt LEMMA}''. At this point of the discussion it is relevant to stress to attendants the big gap between intuition and formalization, letting clear why the proofs of \textit{lemmas} {\tt power\_add\_aux} and  {\tt power\_add\_aux2} require much too additional effort; indeed, they require careful applications of mathematical induction on the exponents. These proofs are proposed to the attendants as further exercise or homework on induction. 

\begin{verbatim}
power_add_aux: CONJECTURE                        
 FORALL (G: (group?), y: (G), n,m: nat): ^(y,n) * ^(y,m) = ^(y, n+m)

power_add_aux2: CONJECTURE                
 FORALL (G: (group?), y: (G), n, m: nat): ^(G)(y,m) * ^(G)(inv(G)(y),n)  =
                                          ^(G)(y, m - n)

power_add: CONJECTURE                        
 FORALL (G: (group?), y: (G), i,j: int): ^(G)(y,i) * ^(G)(y,j) = ^(G)(y, i+j)
\end{verbatim}

In PVS, inductive proofs may be performed by application of the proof commands {\tt (induct)} and {\tt (measure-induct)} that build simple and strong induction schemes, respectively. For instance, for the case of the conjecture {\tt power\_add\_aux} above, the application of simple induction on $n$ with the proof command {\tt (induct "n")} generates the expected goals, given below.   
\[\begin{array}{ll}
\mbox{Induction base:} &  \mbox{\tt FORALL(G, y, m) : \myhat(y,0) * \myhat(y,m) = \myhat(y,m)}\\[1mm] 
\mbox{Induction step:} & \mbox{\tt FORALL j :  FORALL(G, y, m) : \myhat(y,j) * \myhat(y,m) = \myhat(y,j+m) IMPLIES} \\
& \hspace{2.45cm}\mbox{\tt FORALL(G, y, m) : \myhat(y,j+1)$  $* \myhat(y,m) = \myhat(y,j+1+m)}
\end{array}\]

Proving these goals require expanding both the definition of the operator \;\myhat\,,  and of the notion of a group and, the second goal requires additionally Skolemization (using PVS commands as {\tt (skolem)} or {\tt (skeep)}), propositional decomposition and adequate instantiation of the induction hypothesis.  

Finally, the conjecture {\tt cyc\_abel} specifies the property that cyclic groups are Abelian. 

\begin{verbatim}
 cyc_abel: CONJECTURE  cyclic?(G) IMPLIES abelian_group?(G)
\end{verbatim}

Notice that the properties {\tt cyclic?} and then {\tt abelian\_group?} hold just for groups, then one have to prove only that from {\tt cyclic?(G)} one can infer {\tt commutative?[(G)](*)}.   This is obtained by using {\tt power\_add} and the commutativity of integers (indeed, of real numbers regarding addition to prove that $y ^ {i + j} = y ^ {j + i}$) that are provided in the prelude of PVS. 

The theory {\tt pred\_algebra} includes also torsion groups specified as below. 
\begin{verbatim}
torsion?(G): bool = group?(G) AND FORALL (y:(G)): EXISTS (n: posnat) : y^n = e
\end{verbatim}

A thought-provoking example is a conjecture that finite groups are torsion groups. 
\begin{verbatim}
   finite_torsion: CONJECTURE                        
      FORALL (G: (group?)): is_finite(G) IMPLIES torsion?(G)
      \end{verbatim}
      
This is proved showing that for any $y$ element of {\tt G}, the function  $f:{\tt below[N+1]} \rightarrow \{y^0, \ldots, y^{N-1}\}$, where $N$ is the cardinality of {\tt G}, cannot be injective. Here, {\tt below[N+1]} is the subtype of naturals $\{0,\ldots, N\}$. 
Indeed, if $f$ were not injective, there would exist $i<j\leq N$ such that $y^i = y^j$, from which using properties of the operator {\tt \myhat} one concludes that $y^{j-i} = e$.  Non injectivity of $f$ is proved assuming that it is injective. 
In PVS, the predicate {\tt is\_finite(G)} is characterized by the existence of an injective function $g : {\tt (G)} \rightarrow {\tt below[N]}$. The contradiction is obtained since composition of injective functions is injective (formalized in the lemma {\tt composition\_injective} of the PVS prelude library) that would imply that $g\circ f : {\tt below[N+1]} \rightarrow {\tt below[N]}$ is injective.  
      
The last example is the conjecture that for any finite group {\tt G} and element {\tt y} in it, the set of natural powers of {\tt y}  ({\tt power\_gen(y)}) is indeed a subgroup of {\tt G}, specified below.        
\begin{verbatim}
power_gen(x:T): set[T] = { y: T | EXISTS (n:nat) : y = x^n }
           
subgroup?(H:set[T],G:(group?)): bool = subset?(H,G) AND group?(H)

power_gen_fin_group_is_subgroup: CONJECTURE
    FORALL (G: (group?),y:(G)): is_finite(G) IMPLIES subgroup?(power_gen(y),G)
\end{verbatim}

The proof consists in showing that the set {\tt power\_gen(y)}, for $y$ in {\tt G} is a subset of {\tt G} and satisfies all properties of groups. The interesting part of the proof is showing the existence of inverses: for any natural $n$ there exists a natural $m$ such that $y^n * y^m = y^m * y^n = {\tt e}$.  From the previous conjecture ({\tt finite\_torsion}) {\tt G} is a torsion group, which implies that there exists a non-zero natural $k$ such that $(y^n)^k = e$. This allows the choice of $m$ as $n(k-1)$ and with this choice and properties of power one concludes that $y^n * y^{n(k-1)} = y^{n(k-1)} * y^n = y^{nk}$.   An additional property of power, {\tt power\_mult\_aux}, given below, would be required to prove that $y^{nk} = (y^n)^k$ and conclude. 
      
\begin{verbatim}
power_mult_aux: CONJECTURE                   
      FORALL (G:(group?), y:(G), m,n: nat): (y^n)^m  = y^(n*m)
\end{verbatim}
      
\subsection{Formalizations About Symmetric Groups}
More elaborated examples should be discussed to make evident to the attendants that ITPs can be indeed applied to perform serious mathematical proofs, which is not at all a consensus for the mathematicians.  Thus, in addition to the previous basic examples, the tutorial contains additional and more elaborated results. We illustrate this with the properties of a concrete structure that satisfies the definition of groups: symmetric groups. It is formalized in the theory {\tt symmetric[n]} included below, where  the parameter {\tt n} is the  degree of the symmetric group that is a positive natural number. 

\begin{verbatim}
symmetric[n:posnat]: THEORY
  BEGIN
 
  IMPORTING pred_algebra[ [below[n] -> below[n]], o, LAMBDA(i: below[n]): i],
            sets_aux@set_of_functions[below[n],below[n]]

  symmetric: set[[below[n] -> below[n]]] = 
               {f : [below[n] -> below[n]] | bijective?(f)}

  group_symmetric: CONJECTURE group?(symmetric)

  sym_gt2_notabelian: CONJECTURE   
        n >=3 IMPLIES NOT abelian_group?(symmetric)

  sym_cyc_lt2: CONJECTURE cyclic?(symmetric) IMPLIES n < 3

  symmetric_is_finite: CONJECTURE is_finite(symmetric)

  symmetric_is_torsion: CONJECTURE torsion?(symmetric)

  power_gen_subgroup_sym: CONJECTURE 
       FORALL (y:(symmetric)): subgroup?(power_gen(y),symmetric)
        
  END symmetric
\end{verbatim}

The first interesting aspect of this formalization is importing the theory {\tt pred\_algebra} whose parameters were given by the triple {\tt [T, *:[T,T -> T], e:T]}, a non-interpreted non-empty type {\tt T}, a binary operator {\tt *} and a constant {\tt e} of type {\tt T}, and where groups were specified (see Subsection \ref{subsec:simple_deduction}).  This theory is imported by {\tt symmetric[n]} with arguments 
\begin{verbatim}
                [ [below[n] -> below[n]],  o,  LAMBDA(i: below[n]): i ]
\end{verbatim}
Through this importation, the type {\tt T} is interpreted as the set of functions 
from $\{0,\ldots, n-1\}$ into $\{0,\ldots, n-1\}$, the binary operator {\tt *}, as the composition of functions {\tt o}, and the constant {\tt e}, as the identity function,  {\tt LAMBDA(i: below[n]): i}. 

Thus, the set {\tt symmetric} is specified as the subset of  all bijective functions on {\tt below[n]}.  And then the first interesting task is to prove the conjecture {\tt group\_symmetric} that states that {\tt symmetric} is indeed a group: {\tt group?(symmetric)}.  The proof of {\tt group\_symmetric} is, mainly, based on the PVS prelude theory for functions since it provides results such as \emph{``Compositions of bijective functions are also bijective functions"}, \emph{``Bijective functions have  bijective inverses"}, \emph{``Function composition is associative"} and \emph{``The identity function is bijective"}.  

A well-known property about symmetric groups with a degree greater than 2 is that they are not Abelian (Conjecture {\tt sym\_gt2\_notabelian}). The proof of this result consists in providing explicitly two elements of {\tt symmetric} that do not commute regarding the binary operator {\tt o}. For instance, one can construct as witnesses the functions $f$ that maps $0$ into $1$, $1$ into $0$ and fixes all other elements of the domain, and $g$ that maps $0$ into $2$, $2$ into $0$ and fixes all other elements of the domain. Constructive proofs of existential formulas are provided in such a manner.  The above-mentioned functions $f$ and $g$ can be  specified in PVS, respectively, as
\begin{verbatim}
     LAMBDA (i:below[n]) : IF i>=2 THEN i ELSIF i = 0 THEN 1 ELSE 0 ENDIF
     LAMBDA (i:below[n]) : IF i>=3 THEN i ELSIF i = 0 THEN 2 
                           ELSIF i = 1 THEN 1 ELSE 0 ENDIF
\end{verbatim}

Using these functions the proof resumes to show that $f\circ g\neq g\circ f$, which can be done using the fact that $(f\circ g)(0) = 2$ and $(g\circ f)(0) = 1$.

The Conjecture {\tt sym\_cyc\_lt2} is proved as corollary of {\tt sym\_gt2\_notabelian} because  cyclic groups are Abelian.

An interesting formalization, that may illustrate to the attendants how much mathematicians (they!) appeal to their intuition when they develop pen-and-paper proofs is the one of the next conjecture in the theory {\tt symmetric[n]}: {\tt symmetric\_is\_finite}. In general, they justify the finiteness of the symmetric group of degree $n$ just through the observation that the set of bijective functions correspond to the set of $n!$ permutations on a set of $n$ elements,  without worrying about building a proof of this fact.  Without concrete proof of this fact, further complete formalizations would be impossible (such as the fact that symmetric groups are torsion groups). Since in  PVS the predicate {\tt is\_finite(S)} is characterized by the existence of a natural number {\tt N} and an injective function {\tt f:(S)->below[N]}, to prove that the group symmetric is finite, it is necessary to make explicit such an injective function for some adequate {\tt N}.  This formalization is obtained by application of a lemma in the imported theory {\tt sets\_aux@set\_of\_functions} of the NASA PVS library which states that the cardinality of the set of functions from a finite domain to a finite co-domain of respective cardinalities $m$ and $n$ is $n^m$. This is specified as the existence of a bijective function, say {\tt h}, from this set of functions to the type {\tt below[n\myhat m]}. The formalization is concluded proving that the restriction of {\tt h} to the domain {\tt symmetric} is an injective function, which implies the finiteness of {\tt symmetric}. 

As a corollary of the previous result and the fact that finite groups are torsion groups  (see conjecture {\tt finite\_torsion} in the end of Subsection \ref{subsec:simple_deduction}) one obtains the formalization of the conjecture {\tt symmetric\_is\_torsion}.  All other properties of finite groups are then inherited by symmetric groups, such as the fact that  {\tt power\_gen(f)}, for any bijective function {\tt f}  on {\tt below[n]} (see Subsection \ref{subsec:simple_deduction}) is a subgroup of  {\tt symmetric}.

\subsection{Information About Formal Developments}

Some information should be provided to attendants about serious
mathematical important formal developments.  Restricting our attention
just to algebraic theorems, we can mention, for instance, the NASA PVS
Formal Developments library contains general results on algebra
\cite{PVSalgebra} as well advanced results on group theory such as the
Sylow's Theorems \cite{PVSgroups}, and Isomorphism Theorems for groups
and rings, and the Chinese Remainder Theorem for Non-commutative
rings~\cite{LGAA2020}, among others.  Also, results about groups,
rings, and ordered fields are formalized in Coq as part of the FTA
project \cite{geuverscoq2002}. Other important formalizations in Coq
deal with finite group theory \cite{Gonthier2007} culminating in the
formalization of the Feit and Thompson's proof of the Odd Order
Theorem that states that every finite group of odd order is
solvable~\cite{Gonthier13}. Also in Coq, formalizations of real
ordered fields \cite{Cohen2012} and finite fields
\cite{Philipoom2018}, and formalization of rings with explicit
divisibility \cite{cohencoq16} are available. Formalizations of the
Binomial Theorem for rings are available in Nuprl \cite{jackson1995}
and Mizar \cite{cs01}. In ACL2 a hierarchy of algebraic structures
ranging from setoids to vector spaces is built focusing on the
formalization of computer algebra systems \cite{heras15}. The Algebra
Library of Isabelle/HOL \cite{ballarinizabelle2019} provides a wide
range of theorems on mathematical structures, including results on
rings, groups, factorization over ideals, rings of integers and
polynomial rings.  A formalization of the First Isomorphism Theorem
for rings is also available in Mizar \cite{Kornilowicz2014}.

\section{Discussion}\label{sec:discussion}

The main interest of this position paper is to show how ITPs can be
promoted among users of related areas, who do not necessarily require
or desire to develop a strong background in proof theory and
mathematical deduction, but who just need or want to apply these tools
in their areas of expertise.  Usually, we can reach the attention of
mathematicians in workshops in which their time availability is
restricted and for which short-courses and tutorials should be limited
to a few hours (usually, from two to at most eight hours). Because of
this, our proposal requires a huge effort to adopt nice well-known
mathematical examples that fulfil the real interests of the
attendants, it focuses on the mathematical aspects of the proofs and
omits long dissertations about logical systems, deduction and proof
theory. Also, our position is that good graduation and diversity of
the complexity of the selected examples contributes to illustrate
clearly to the attendants the real power of ITPs to prove elaborated
mathematical theorems.  This contrast with the high flexibility we
have when teaching logical deduction to CS students in our
one-semester sixty-four-hour course on Computational Logic, mentioned
in the introduction, for which we have enough time and can focus on
natural and sequent calculus-based deduction and spend at least a
third of the semester training students in the application of
deductive tools (\cite{AR_FMTeaching2019,ayala2017applied}).

\bibliographystyle{eptcs}
\bibliography{Revised_ThEduIP4Mah}

\begin{thebibliography}{10}
\providecommand{\bibitemdeclare}[2]{}
\providecommand{\surnamestart}{}
\providecommand{\surnameend}{}
\providecommand{\urlprefix}{Available at }
\providecommand{\url}[1]{\texttt{#1}}
\providecommand{\href}[2]{\texttt{#2}}
\providecommand{\urlalt}[2]{\href{#1}{#2}}
\providecommand{\doi}[1]{doi:\urlalt{http://dx.doi.org/#1}{#1}}
\providecommand{\bibinfo}[2]{#2}

\bibitemdeclare{book}{theBook2018}
\bibitem{theBook2018}
\bibinfo{author}{Martin \surnamestart Aigner\surnameend} \&
  \bibinfo{author}{G\"{u}nter~M. \surnamestart Ziegler\surnameend}
  (\bibinfo{year}{2018}): \emph{\bibinfo{title}{{Proofs from THE BOOK}}},
  \bibinfo{edition}{6th} edition.
\newblock \bibinfo{publisher}{Springer}, \doi{10.1007/978-3-662-57265-8}.

\bibitemdeclare{inproceedings}{AR_FMTeaching2019}
\bibitem{AR_FMTeaching2019}
\bibinfo{author}{Ariane~Alves \surnamestart Almeida\surnameend},
  \bibinfo{author}{Ana~Cristina \surnamestart Rocha-Oliveira\surnameend},
  \bibinfo{author}{Thiago~Mendon\c{c}a \surnamestart
  Ferreira~Ramos\surnameend}, \bibinfo{author}{Fl\'avio Leonardo~C.
  \surnamestart de~Moura\surnameend} \& \bibinfo{author}{Mauricio \surnamestart
  Ayala-Rinc\'on\surnameend} (\bibinfo{year}{2019}): \emph{\bibinfo{title}{{The
  Computational Relevance of Formal Logic Through Formal Proofs}}}.
\newblock In: {\sl \bibinfo{booktitle}{Proceedings 3rd Formal Methods Teaching
  {FMTea}}}, {\sl \bibinfo{series}{LNCS}} \bibinfo{volume}{11758},
  \bibinfo{publisher}{Springer}, pp. \bibinfo{pages}{81--96},
  \doi{10.1007/978-3-030-32441-4\_6}.

\bibitemdeclare{techreport}{ballarinizabelle2019}
\bibitem{ballarinizabelle2019}
\bibinfo{author}{Jes\'{u}s \surnamestart Aransay\surnameend},
  \bibinfo{author}{Clemens \surnamestart Ballarin\surnameend},
  \bibinfo{author}{Martin \surnamestart Baillon\surnameend},
  \bibinfo{author}{Paulo~Em\'{i}lio \surnamestart de~Vilhena\surnameend},
  \bibinfo{author}{Stephan \surnamestart Hohe\surnameend},
  \bibinfo{author}{Florian \surnamestart Kamm\"{u}ller\surnameend} \&
  \bibinfo{author}{Lawrence~C. \surnamestart Paulson\surnameend}
  (\bibinfo{year}{2019}): \emph{\bibinfo{title}{{The Isabelle/HOL Algebra
  Library}}}.
\newblock \bibinfo{type}{Technical Report}, \bibinfo{institution}{Isabelle
  Library, University of Cambridge Computer Laboratory and Technische
  Universit\"at M\"unchen}.
\newblock
  \urlprefix\url{https://isabelle.in.tum.de/dist/library/HOL/HOL-Algebra/document.pdf}.

\bibitemdeclare{book}{ayala2017applied}
\bibitem{ayala2017applied}
\bibinfo{author}{Mauricio \surnamestart Ayala-Rinc{\'o}n\surnameend} \&
  \bibinfo{author}{Fl\'avio Leonardo~C. \surnamestart de~Moura\surnameend}
  (\bibinfo{year}{2017}): \emph{\bibinfo{title}{{Applied Logic for Computer
  Scientists: Computational Deduction and Formal Proofs}}}.
\newblock \bibinfo{series}{{UTiCS}}, \bibinfo{publisher}{Springer},
  \doi{10.1007/978-3-319-51653-0}.

\bibitemdeclare{article}{BaazHLRS08}
\bibitem{BaazHLRS08}
\bibinfo{author}{Matthias \surnamestart Baaz\surnameend},
  \bibinfo{author}{Stefan \surnamestart Hetzl\surnameend},
  \bibinfo{author}{Alexander \surnamestart Leitsch\surnameend},
  \bibinfo{author}{Clemens \surnamestart Richter\surnameend} \&
  \bibinfo{author}{Hendrik \surnamestart Spohr\surnameend}
  (\bibinfo{year}{2008}): \emph{\bibinfo{title}{{CERES:} An analysis of
  F{\"{u}}rstenberg's proof of the infinity of primes}}.
\newblock {\sl \bibinfo{journal}{Theor. Comput. Sci.}}
  \bibinfo{volume}{403}(\bibinfo{number}{2-3}), pp. \bibinfo{pages}{160--175},
  \doi{10.1016/j.tcs.2008.02.043}.

\bibitemdeclare{inbook}{Bayer2019}
\bibitem{Bayer2019}
\bibinfo{author}{Jonas \surnamestart Bayer\surnameend}, \bibinfo{author}{Marco
  \surnamestart David\surnameend}, \bibinfo{author}{Abhik \surnamestart
  Pal\surnameend} \& \bibinfo{author}{Benedikt \surnamestart Stock\surnameend}
  (\bibinfo{year}{2019}): \emph{\bibinfo{title}{{Beginners’ Quest to
  Formalize Mathematics: A Feasibility Study in Isabelle}}}, pp.
  \bibinfo{pages}{16--27}.
\newblock \doi{10.1007/978-3-030-23250-4\_2}.

\bibitemdeclare{inproceedings}{Blazy_FMTea2019}
\bibitem{Blazy_FMTea2019}
\bibinfo{author}{Sandrine \surnamestart Blazy\surnameend}
  (\bibinfo{year}{2019}): \emph{\bibinfo{title}{{Teaching Deductive
  Verification in Why3 to Undergraduate Students}}}.
\newblock In: {\sl \bibinfo{booktitle}{Proceedings 3rd Formal Methods Teaching
  {FMTea}}}, {\sl \bibinfo{series}{LNCS}} \bibinfo{volume}{11758},
  \bibinfo{publisher}{Springer}, pp. \bibinfo{pages}{52--66},
  \doi{10.1007/978-3-030-32441-4\_4}.

\bibitemdeclare{misc}{PVSalgebra}
\bibitem{PVSalgebra}
\bibinfo{author}{Ricky \surnamestart Butler\surnameend} \&
  \bibinfo{author}{David \surnamestart Lester\surnameend}
  (\bibinfo{year}{2007}): \emph{\bibinfo{title}{{A PVS \emph{Theory} for
  Abstract Algebra}}}.
\newblock \urlprefix\url{https://github.com/nasa/pvslib/tree/pvs7.0/algebra}.
\newblock \bibinfo{note}{Accessed in June 17, 2020}.

\bibitemdeclare{article}{cohencoq16}
\bibitem{cohencoq16}
\bibinfo{author}{Guillaume \surnamestart Cano\surnameend},
  \bibinfo{author}{Cyril \surnamestart Cohen\surnameend},
  \bibinfo{author}{Maxime \surnamestart D\'{e}n\`{e}s\surnameend},
  \bibinfo{author}{Anders \surnamestart M\"{o}rtberg\surnameend} \&
  \bibinfo{author}{Vincent \surnamestart Siles\surnameend}
  (\bibinfo{year}{2016}): \emph{\bibinfo{title}{{Formalized linear algebra over
  Elementary Divisor Rings in {Coq}}}}.
\newblock {\sl \bibinfo{journal}{Logical Methods in Computer Science}}
  \bibinfo{volume}{12}(\bibinfo{number}{2:7}), pp. \bibinfo{pages}{1--23},
  \doi{10.2168/LMCS-12(2:7)2016}.

\bibitemdeclare{article}{Cohen2012}
\bibitem{Cohen2012}
\bibinfo{author}{Cyril \surnamestart Cohen\surnameend} \&
  \bibinfo{author}{Assia \surnamestart Mahboubi\surnameend}
  (\bibinfo{year}{2012}): \emph{\bibinfo{title}{{Formal proofs in real
  algebraic geometry: from ordered fields to quantifier elimination}}}.
\newblock {\sl \bibinfo{journal}{Logical Methods in Computer Science}}
  \bibinfo{volume}{8}(\bibinfo{number}{1:2}), pp. \bibinfo{pages}{1--40},
  \doi{10.2168/LMCS-8(1:2)2012}.

\bibitemdeclare{inproceedings}{Dubois_FMTea2019}
\bibitem{Dubois_FMTea2019}
\bibinfo{author}{Catherine \surnamestart Dubois\surnameend},
  \bibinfo{author}{Virgile \surnamestart Prevosto\surnameend} \&
  \bibinfo{author}{Guillaume \surnamestart Burel\surnameend}
  (\bibinfo{year}{2019}): \emph{\bibinfo{title}{{Teaching Formal Methods to
  Future Engineers}}}.
\newblock In: {\sl \bibinfo{booktitle}{Proceedings 3rd Formal Methods Teaching
  {FMTea}}}, {\sl \bibinfo{series}{LNCS}} \bibinfo{volume}{11758},
  \bibinfo{publisher}{Springer}, pp. \bibinfo{pages}{69--80},
  \doi{10.1007/978-3-030-32441-4\_5}.

\bibitemdeclare{article}{Eberl20a}
\bibitem{Eberl20a}
\bibinfo{author}{Manuel \surnamestart Eberl\surnameend} (\bibinfo{year}{2020}):
  \emph{\bibinfo{title}{{F\"urstenberg's topology and his proof of the
  infinitude of primes}}}.
\newblock {\sl \bibinfo{journal}{Arch. Formal Proofs}} \bibinfo{volume}{2020}.
\newblock
  \urlprefix\url{https://www.isa-afp.org/entries/Furstenberg\_Topology.html}.

\bibitemdeclare{book}{Fraleigh}
\bibitem{Fraleigh}
\bibinfo{author}{John~B. \surnamestart Fraleigh\surnameend}
  (\bibinfo{year}{2002}): \emph{\bibinfo{title}{{A First Course in Abstract
  Algebra}}}, \bibinfo{edition}{7th} edition.
\newblock \bibinfo{publisher}{Pearson}.

\bibitemdeclare{article}{Fuerstenberg1955}
\bibitem{Fuerstenberg1955}
\bibinfo{author}{Hillel \surnamestart F\"urstenberg\surnameend}
  (\bibinfo{year}{1955}): \emph{\bibinfo{title}{{On the Infinitude of
  Primes}}}.
\newblock {\sl \bibinfo{journal}{Amer. Math, Monthly}}
  \bibinfo{volume}{62}(\bibinfo{number}{5}), p. \bibinfo{pages}{353},
  \doi{10.2307/2307043}.

\bibitemdeclare{misc}{PVSgroups}
\bibitem{PVSgroups}
\bibinfo{author}{Andr{\'{e}}~Luiz \surnamestart Galdino\surnameend}
  (\bibinfo{year}{2011}): \emph{\bibinfo{title}{{A PVS \emph{Theory} for
  Groups}}}.
\newblock \urlprefix\url{https://github.com/nasa/pvslib/tree/pvs7.0/groups}.
\newblock \bibinfo{note}{Accessed in June 17, 2020}.

\bibitemdeclare{article}{geuverscoq2002}
\bibitem{geuverscoq2002}
\bibinfo{author}{Herman \surnamestart Geuvers\surnameend},
  \bibinfo{author}{Randy \surnamestart Pollack\surnameend},
  \bibinfo{author}{Freek \surnamestart Wiedijk\surnameend} \&
  \bibinfo{author}{Jan \surnamestart Zwanenburg\surnameend}
  (\bibinfo{year}{2002}): \emph{\bibinfo{title}{A Constructive Algebraic
  Hierarchy in {Coq}}}.
\newblock {\sl \bibinfo{journal}{Journal of Symbolic Computation}}
  \bibinfo{volume}{34}(\bibinfo{number}{4}), pp. \bibinfo{pages}{271--286},
  \doi{10.1006/jsco.2002.0552}.

\bibitemdeclare{inproceedings}{Gonthier13}
\bibitem{Gonthier13}
\bibinfo{author}{Georges \surnamestart Gonthier\surnameend},
  \bibinfo{author}{Andrea \surnamestart Asperti\surnameend},
  \bibinfo{author}{Jeremy \surnamestart Avigad\surnameend},
  \bibinfo{author}{Yves \surnamestart Bertot\surnameend},
  \bibinfo{author}{Cyril \surnamestart Cohen\surnameend},
  \bibinfo{author}{Fran{\c{c}}ois \surnamestart Garillot\surnameend},
  \bibinfo{author}{St{\'{e}}phane~Le \surnamestart Roux\surnameend},
  \bibinfo{author}{Assia \surnamestart Mahboubi\surnameend},
  \bibinfo{author}{Russell \surnamestart O'Connor\surnameend},
  \bibinfo{author}{Sidi~Ould \surnamestart Biha\surnameend},
  \bibinfo{author}{Ioana \surnamestart Pasca\surnameend},
  \bibinfo{author}{Laurence \surnamestart Rideau\surnameend},
  \bibinfo{author}{Alexey \surnamestart Solovyev\surnameend},
  \bibinfo{author}{Enrico \surnamestart Tassi\surnameend} \&
  \bibinfo{author}{Laurent \surnamestart Th{\'{e}}ry\surnameend}
  (\bibinfo{year}{2013}): \emph{\bibinfo{title}{{A Machine-Checked Proof of the
  Odd Order Theorem}}}.
\newblock In: {\sl \bibinfo{booktitle}{4th International Conference on
  Interactive Theorem Proving {ITP}}}, {\sl \bibinfo{series}{LNCS}}
  \bibinfo{volume}{7998}, \bibinfo{publisher}{Springer}, pp.
  \bibinfo{pages}{163--179}, \doi{10.1007/978-3-642-39634-2\_14}.

\bibitemdeclare{inproceedings}{Gonthier2007}
\bibitem{Gonthier2007}
\bibinfo{author}{Georges \surnamestart Gonthier\surnameend},
  \bibinfo{author}{Assia \surnamestart Mahboubi\surnameend},
  \bibinfo{author}{Laurence \surnamestart Rideau\surnameend},
  \bibinfo{author}{Enrico \surnamestart Tassi\surnameend} \&
  \bibinfo{author}{Laurent \surnamestart Th{\'{e}}ry\surnameend}
  (\bibinfo{year}{2007}): \emph{\bibinfo{title}{{A Modular Formalisation of
  Finite Group Theory}}}.
\newblock In: {\sl \bibinfo{booktitle}{20th International Conference Theorem
  Proving in Higher Order Logics {TPHOLs}}}, {\sl \bibinfo{series}{LNCS}}
  \bibinfo{volume}{4732}, \bibinfo{publisher}{Springer}, pp.
  \bibinfo{pages}{86--101}, \doi{10.1007/978-3-540-74591-4\_8}.

\bibitemdeclare{article}{Harrison11}
\bibitem{Harrison11}
\bibinfo{author}{John \surnamestart Harrison\surnameend}
  (\bibinfo{year}{2011}): \emph{\bibinfo{title}{{A formal proof of Pick's
  Theorem}}}.
\newblock {\sl \bibinfo{journal}{Math. Struct. Comput. Sci.}}
  \bibinfo{volume}{21}(\bibinfo{number}{4}), pp. \bibinfo{pages}{715--729},
  \doi{10.1017/S0960129511000089}.

\bibitemdeclare{article}{heras15}
\bibitem{heras15}
\bibinfo{author}{J{\'o}nathan \surnamestart Heras\surnameend},
  \bibinfo{author}{Francisco~Jes{\'u}s \surnamestart
  Mart{\'i}n-Mateos\surnameend} \& \bibinfo{author}{Vico \surnamestart
  Pascual\surnameend} (\bibinfo{year}{2015}): \emph{\bibinfo{title}{{Modelling
  algebraic structures and morphisms in {ACL2}}}}.
\newblock {\sl \bibinfo{journal}{Applicable Algebra in Engineering,
  Communication and Computing}} \bibinfo{volume}{26}(\bibinfo{number}{3}), pp.
  \bibinfo{pages}{277--303}, \doi{10.1007/s00200-015-0252-9}.

\bibitemdeclare{book}{herstein75}
\bibitem{herstein75}
\bibinfo{author}{Israel (Yitzchak)~Nathan \surnamestart Herstein\surnameend}
  (\bibinfo{year}{1975}): \emph{\bibinfo{title}{Topics in algebra}},
  \bibinfo{edition}{2nd} edition.
\newblock \bibinfo{publisher}{John Wiley \& Sons}.

\bibitemdeclare{book}{hungerford80}
\bibitem{hungerford80}
\bibinfo{author}{Thomas~W. \surnamestart Hungerford\surnameend}
  (\bibinfo{year}{1980}): \emph{\bibinfo{title}{Algebra}}.
\newblock {\sl \bibinfo{series}{Graduate Texts in
  Mathematics}}~\bibinfo{volume}{73}, \bibinfo{publisher}{Springer},
  \doi{10.1007/978-1-4612-6101-8}.
\newblock \bibinfo{note}{Reprint of the 1974 original}.

\bibitemdeclare{phdthesis}{jackson1995}
\bibitem{jackson1995}
\bibinfo{author}{Paul~Bernard \surnamestart Jackson\surnameend}
  (\bibinfo{year}{1995}): \emph{\bibinfo{title}{{Enhancing the Nuprl Proof
  Development System and Applying it to Computational Abstract Algebra}}}.
\newblock Ph.D. thesis, \bibinfo{school}{Cornell University}.

\bibitemdeclare{article}{Kornilowicz2014}
\bibitem{Kornilowicz2014}
\bibinfo{author}{Artur \surnamestart Kornilowicz\surnameend} \&
  \bibinfo{author}{Christoph \surnamestart Schwarzweller\surnameend}
  (\bibinfo{year}{2014}): \emph{\bibinfo{title}{{The First Isomorphism Theorem
  and Other Properties of Rings}}}.
\newblock {\sl \bibinfo{journal}{Formalized Mathematics}}
  \bibinfo{volume}{22}(\bibinfo{number}{4}), pp. \bibinfo{pages}{291-- 301},
  \doi{10.2478/forma-2014-0029}.

\bibitemdeclare{unpublished}{LGAA2020}
\bibitem{LGAA2020}
\bibinfo{author}{Thaynara~Arielly \surnamestart de~Lima\surnameend},
  \bibinfo{author}{Andr{\'{e}}~Luiz \surnamestart Galdino\surnameend},
  \bibinfo{author}{Andr{\'{e}}ia~B. \surnamestart Avelar\surnameend} \&
  \bibinfo{author}{Mauricio \surnamestart Ayala-Rinc{\'{o}}n\surnameend}
  (\bibinfo{year}{2020}): \emph{\bibinfo{title}{Formalization of Ring Theory in
  PVS - Isomorphism Theorems, Principal, Prime and Maximal Ideals,Chinese
  Remainder Theorem}}.
\newblock \urlprefix\url{http://ayala.mat.unb.br/publications.html}.

\bibitemdeclare{mastersthesis}{Philipoom2018}
\bibitem{Philipoom2018}
\bibinfo{author}{Jade \surnamestart Philipoom\surnameend}
  (\bibinfo{year}{2018}): \emph{\bibinfo{title}{{Correct-by-Construction Finite
  Field Arithmetic in Coq}}}.
\newblock Master's thesis, \bibinfo{school}{Master of Engineering in Computer
  Science, MIT}.

\bibitemdeclare{inproceedings}{Rozier_FMTea2019}
\bibitem{Rozier_FMTea2019}
\bibinfo{author}{Kristin~Yvonne \surnamestart Rozier\surnameend}
  (\bibinfo{year}{2019}): \emph{\bibinfo{title}{{On Teaching Applied Formal
  Methods in Aerospace Engineering}}}.
\newblock In: {\sl \bibinfo{booktitle}{Proceedings 3rd Formal Methods Teaching
  {FMTea}}}, {\sl \bibinfo{series}{LNCS}} \bibinfo{volume}{11758},
  \bibinfo{publisher}{Springer}, pp. \bibinfo{pages}{111--131},
  \doi{10.1007/978-3-030-32441-4\_8}.

\bibitemdeclare{article}{cs01}
\bibitem{cs01}
\bibinfo{author}{Christoph \surnamestart Schwarzweller\surnameend}
  (\bibinfo{year}{2003}): \emph{\bibinfo{title}{{The Binomial Theorem for
  Algebraic Structures}}}.
\newblock {\sl \bibinfo{journal}{Journal of Formalized Mathematics}}
  \bibinfo{volume}{12}(\bibinfo{number}{3}), pp. \bibinfo{pages}{559--564}.
\newblock \urlprefix\url{http://mizar.org/JFM/Vol12/binom.html}.

\bibitemdeclare{inproceedings}{SilvaLG18}
\bibitem{SilvaLG18}
\bibinfo{author}{Andr{\'{e}}ia B.~Avelar \surnamestart da~Silva\surnameend},
  \bibinfo{author}{Thaynara~Arielly \surnamestart de~Lima\surnameend} \&
  \bibinfo{author}{Andr{\'{e}}~Luiz \surnamestart Galdino\surnameend}
  (\bibinfo{year}{2018}): \emph{\bibinfo{title}{Formalizing Ring Theory in
  {PVS}}}.
\newblock In: {\sl \bibinfo{booktitle}{9th International Conference on
  Interactive Theorem Proving {ITP}}}, {\sl \bibinfo{series}{LNCS}}
  \bibinfo{volume}{10895}, \bibinfo{publisher}{Springer}, pp.
  \bibinfo{pages}{40--47}, \doi{10.1007/978-3-319-94821-8\_3}.

\end{thebibliography}
\end{document}